# The zero exemplar distance problem[*]

Minghui Jiang

Department of Computer Science, Utah State University, Logan, UT 84322, USA
`mjiang@cc.usu.edu`

November 5, 2010

## Abstract

Given two genomes with duplicate genes, ZERO EXEMPLAR DISTANCE is the problem of deciding whether the two genomes can be reduced to the same genome without duplicate genes by deleting all but one copy of each gene in each genome. Blin, Fertin, Sikora, and Vialette recently proved that ZERO EXEMPLAR DISTANCE for monochromosomal genomes is NP-hard even if each gene appears at most two times in each genome, thereby settling an important open question on genome rearrangement in the exemplar model. In this paper, we give a very simple alternative proof of this result. We also study the problem ZERO EXEMPLAR DISTANCE for multichromosomal genomes without gene order, and prove the analogous result that it is also NP-hard even if each gene appears at most two times in each genome. For the positive direction, we show that both variants of ZERO EXEMPLAR DISTANCE admit polynomial-time algorithms if each gene appears exactly once in one genome and at least once in the other genome. In addition, we present a polynomial-time algorithm for the related problem EXEMPLAR LONGEST COMMON SUBSEQUENCE in the special case that each mandatory symbol appears exactly once in one input sequence and at least once in the other input sequence. This answers an open question of Bonizzoni et al. We also show that ZERO EXEMPLAR DISTANCE for multichromosomal genomes without gene order is fixed-parameter tractable if the parameter is the maximum number of chromosomes in each genome.

## 1 Introduction

Given two genomes with duplicate genes, GENOME REARRANGEMENT WITH GENE FAMILIES [12] is the problem of deleting all but one copy of each gene in each genome, so as to minimize some rearrangement distance between the two reduced genomes. The minimum rearrangement distance thus attained is called the *exemplar distance* between the two genomes. For example, each of the following two monochromosomal genomes

$$G_1 : \quad -4 \ +1 \ +2 \ +3 \ -5 \ +1 \ +2 \ +3 \ -6$$
$$G_2 : \quad -1 \ -4 \ +1 \ +2 \ -5 \ +3 \ -2 \ -6 \ +3$$

---

[*]Supported in part by NSF grant DBI-0743670. A preliminary version of this paper (including Theorem 1, Theorem 2, and a weaker version of Theorem 4) appeared in Proceedings of the 8th Annual RECOMB Satellite Workshop on Comparative Genomics (RECOMB-CG 2010) [11].

has at most two copies of each gene, and each of the following two reduced genomes

$$G'_1: \quad -4 \ +1 \ +2 \ -5 \ +3 \ -6$$
$$G'_2: \quad -4 \ +1 \ +2 \ -5 \ +3 \ -6$$

has exactly one copy of each gene. Recall that in the study of genome rearrangement, a *gene* is usually represented by a signed integer: the absolute value of the integer (the unsigned integer) denotes the gene family to which the gene belongs; the sign of the integer denotes the orientation of the gene in its chromosome. Then a *chromosome* is a sequence of signed integers, and a *genome* is a collection of chromosomes.

GENOME REARRANGEMENT WITH GENE FAMILIES is not a single problem but a whole class of related problems, because the choice of rearrangement distance is not unique. This choice becomes irrelevant, however, when we ask the fundamental question: *Is the distance zero?* In the example above, the two reduced genomes $G'_1$ and $G'_2$ are identical, thus the exemplar distance between the two original genomes $G_1$ and $G_2$ is zero for any reasonable choice of rearrangement distance.

In this paper, we study the most basic version of the problem GENOME REARRANGEMENT WITH GENE FAMILIES: Given two sequences of signed integers, ZERO EXEMPLAR DISTANCE (for monochromosomal genomes) is the problem of deciding whether the two sequences have a common subsequence including each unsigned integer exactly once in either positive or negative form.

Due to its generic nature, the problem ZERO EXEMPLAR DISTANCE has been extensively studied by several groups of researchers [5, 4, 2] focusing on different rearrangement distances, and, not surprisingly, has acquired several different names. Except for trivial distinctions, ZERO EXEMPLAR DISTANCE is essentially the same problem as ZERO EXEMPLAR CONSERVED INTERVAL DISTANCE [5], EXEMPLAR LONGEST COMMON SUBSEQUENCE (deciding whether a feasible solution exists) [4], and ZERO EXEMPLAR BREAKPOINT DISTANCE [2].

It is easy to check that if only one of the two genomes has duplicate genes, then ZERO EXEMPLAR DISTANCE can be solved in linear time: we simply need to decide whether the genome without duplicates is a subsequence of the genome with duplicates. In sharp contrast, if both genomes contain duplicate genes, then even if each gene appears at most three times in each genome, the problem ZERO EXEMPLAR DISTANCE is already NP-hard, as shown independently in three papers [5, 4, 2]. The quest for the exact boundary between polynomial solvability and NP-hardness led to the following open question first raised by Chen et al. in 2006:

**Question 1** (Chen, Fowler, Fu, and Zhu, 2006 [5]). *Is the problem ZERO EXEMPLAR DISTANCE for monochromosomal genomes still NP-hard if each gene appears at most two times in each genome?*

This question was finally settled in the affirmative by Blin et al. in 2009:

**Theorem 1** (Blin, Fertin, Sikora, and Vialette, 2009 [3]). ZERO EXEMPLAR DISTANCE *for monochromosomal genomes is NP-hard even if each gene appears at most two times in each genome.*

In Section 2, we give a very simple alternative proof of this theorem.

Both the previous proof of Theorem 1 [3] and our alternative proof depend crucially on the order of the genes in the chromosomes. One may naturally wonder whether the complexity of ZERO EXEMPLAR DISTANCE would change if gene order is not known. Note that genome rearrangement distances such as the syntenic distance [8] can be defined in the absence of gene order.

Now model each chromosome as a set of unsigned integers instead of a sequence of signed integers. Then ZERO EXEMPLAR DISTANCE for multichromosomal genomes without gene order is the following problem: Given two collections $G_1$ and $G_2$ of subsets of the same ground set $S$ of unsigned integers, decide whether both $G_1$ and $G_2$ can be reduced, by deleting elements from subsets and deleting subsets from collections, to the same collection $G'$ of subsets of $S$ such that each unsigned integer in $S$ is contained in exactly one subset in $G'$, i.e., $G'$ is a partition of $S$. For example,

$$\begin{aligned} S: \quad & \{1,2,3,4,5\} \\ G_1: \quad & \{1,2,3\} \quad \{2,3,4\} \quad \{4,5\} \\ G_2: \quad & \{1,2\} \quad \{2,3,4\} \quad \{3,4,5\} \quad \{1,5\} \\ G': \quad & \{1,2\} \quad \{3\} \quad \{4,5\} \end{aligned}$$

In Section 3, we prove the following theorem analogous to Theorem 1:

**Theorem 2.** ZERO EXEMPLAR DISTANCE *for multichromosomal genomes without gene order is NP-hard even if each gene appears at most two times in each genome.*

As decision problems, both variants of ZERO EXEMPLAR DISTANCE, for monochromosomal genomes and for multichromosomal genomes without gene order, are in NP. Thus, following the NP-hardness results in Theorem 1 and Theorem 2, these two decision problems are both NP-complete. Moreover, the NP-hardness results in Theorem 1 and Theorem 2 imply that unless NP = P, the corresponding minimization problems of computing the exemplar distance between two genomes do not admit *any* approximation. We refer to [5, 6, 4, 2, 1] for related results.

The problem ZERO EXEMPLAR DISTANCE for monochromosomal genomes, as mentioned earlier, has been studied under several different names. Given two sequences $A$ and $B$ over an alphabet $\Sigma = \Sigma_1 \cup \Sigma_2$, where $\Sigma_1$ is a set of *mandatory* symbols and $\Sigma_2$ is a set of *optional* symbols, EXEMPLAR LONGEST COMMON SUBSEQUENCE [4] is the problem of finding a longest common subsequence of $A$ and $B$ that contains all mandatory symbols in $\Sigma_1$. For example, if $\Sigma_1 = \{1, 2, 3\}$ and $\Sigma_2 = \{4, 5\}$, then $C = 124355$ is an exemplar longest common subsequence of the two sequences $A = 12423545$ and $B = 1142443555$.

Due to the strict requirement on mandatory symbols, EXEMPLAR LONGEST COMMON SUBSEQUENCE does not always have a feasible solution. It is not difficult to see that simply deciding whether a feasible solution to EXEMPLAR LONGEST COMMON SUBSEQUENCE exists for two sequences $A$ and $B$ is the same as the problem ZERO EXEMPLAR DISTANCE for two monochromosomal genomes $A'$ and $B'$ obtained from $A$ and $B$ by deleting all optional symbols. Recall that the problem ZERO EXEMPLAR DISTANCE for monochromosomal genomes becomes trivial when only one of the two genomes has duplicate genes. For the equivalent problem of deciding whether a feasible solution to EXEMPLAR LONGEST COMMON SUBSEQUENCE exists, Bonizzoni et al. [4] showed another tractable special case: If each mandatory symbol appears a total of at most three times in $A$ and $B$, then there is a polynomial-time algorithm, based on 2SAT, that decides whether $A$ and $B$ have a common subsequence containing all mandatory symbols. This algorithm does not solve the maximization problem, however, and the following question was left open:

**Question 2** (Bonizzoni et al. [4]). *Is there a polynomial-time algorithm for* EXEMPLAR LONGEST COMMON SUBSEQUENCE *in the special case that each mandatory symbol appears a total of at most three times in the two input sequences?*

Without loss of generality, we assume that each input sequence contains each symbol in the alphabet at least once. If each mandatory symbol appears a total of at most three times in the two input sequences, then it must appear exactly once in one sequence, and at least once in the other sequence, as in the example shown earlier. In Section 4, we prove the following theorem that complements Theorem 1 and answers the open question of Bonizzoni et al. in the affirmative:

**Theorem 3.** ZERO EXEMPLAR DISTANCE *for monochromosomal genomes admits a polynomial-time algorithm in the special case that each gene appears exactly once in one genome and at least once in the other genome.* EXEMPLAR LONGEST COMMON SUBSEQUENCE *admits a polynomial-time algorithm in the special case that each mandatory symbol appears exactly once in one input sequence and at least once in the other input sequence.*

Finally, in Section 5, we prove the following theorem that complements Theorem 2:

**Theorem 4.** ZERO EXEMPLAR DISTANCE *for multichromosomal genomes without gene order admits a polynomial-time algorithm in the special case that each gene appears exactly once in one genome and at least once in the other genome, and is fixed-parameter tractable if the parameter is the maximum number of chromosomes in each genome.*

## 2 Alternative Proof of Theorem 1

We prove that ZERO EXEMPLAR DISTANCE for monochromosomal genomes is NP-hard by a reduction from the well-known NP-complete problem 3SAT [9]. Let $(V, E)$ be a 3SAT instance, where $V = \{v_1, \ldots, v_n\}$ is a set of $n$ boolean variables, $E = \{e_1, \ldots, e_m\}$ is a conjunctive boolean formula of $m$ clauses, and each clause in $E$ is a disjunction of exactly three literals of the variables in $V$. We will construct two sequences (genomes) $G_1$ and $G_2$ over $2n + 6m + 1$ distinct unsigned integers (genes):

- Two *variable genes* $x_i, y_i$ for each variable $v_i$, $1 \le i \le n$;
- Three *clause genes* $a_j, b_j, c_j$ for each clause $e_j$, $1 \le j \le m$;
- Three *literal genes* $r_j, s_j, t_j$ for the three literals of each clause $e_j$, $1 \le j \le m$;
- One *separator gene* $z$.

In our construction, all genes appear in the positive orientation in the two genomes, so we will omit the signs in our description. The two genomes $G_1$ and $G_2$ are represented schematically as follows:

$$G_1: \quad \langle v_1 \rangle \ldots \langle v_n \rangle \; z \; \langle e_1 \rangle \ldots \langle e_m \rangle$$
$$G_2: \quad \langle v_1 \rangle \ldots \langle v_n \rangle \; z \; \langle e_1 \rangle \ldots \langle e_m \rangle$$

For each variable $v_i$, the variable gadget $\langle v_i \rangle$ consists of one copy of $x_i$ and two copies of $y_i$ in $G_1$, two copies of $x_i$ and one copy of $y_i$ in $G_2$, and, for each literal of the variable in the clauses, one copy of the corresponding literal gene ($r_j$, $s_j$, or $t_j$ for some clause $e_j$) in each genome. Let $p_{i,1}, \ldots, p_{i,k_i}$ be the literal genes for the positive literals of $v_i$, and let $q_{i,1}, \ldots, q_{i,l_i}$ be the literal genes for the negative literals of $v_i$. The genes $x_i, y_i, p_{i,1}, \ldots, p_{i,k_i}, q_{i,1}, \ldots, q_{i,l_i}$ in the variable gadget $\langle v_i \rangle$ are arranged in the following pattern in the two genomes:

$$G_1 \langle v_i \rangle : \quad y_i \; p_{i,1} \ldots p_{i,k_i} \; x_i \; q_{i,1} \ldots q_{i,l_i} \; y_i$$
$$G_2 \langle v_i \rangle : \quad p_{i,1} \ldots p_{i,k_i} \; x_i \; y_i \; x_i \; q_{i,1} \ldots q_{i,l_i}$$

For each clause $e_j$, the clause gadget $\langle e_j \rangle$ consists of two copies of each clause gene $a_j, b_j, c_j$ and one copy of each literal gene $r_j, s_j, t_j$. These genes in $\langle e_j \rangle$ are arranged in the following pattern in the two genomes:

$$G_1 \langle e_j \rangle : \quad r_j\, a_j\, b_j\, c_j\, s_j\, a_j\, b_j\, c_j\, t_j$$
$$G_2 \langle e_j \rangle : \quad a_j\, r_j\, b_j\, a_j\, s_j\, c_j\, b_j\, t_j\, c_j$$

This completes the construction. It is easy to check that each gene appears at most two times in each genome, and that each genome includes exactly $3n + 12m + 1$ genes including duplicates. We give an example:

**Example 1.** *For a 3SAT instance of 4 variables and 2 clauses $e_1 = \{r_1 = v_1, s_1 = \neg v_2, t_1 = \neg v_3\}$ and $e_2 = \{r_2 = \neg v_1, s_2 = v_3, t_2 = v_4\}$, the reduction constructs the following two genomes:*

$$
\begin{aligned}
G_1 : \quad & y_1 r_1 x_1 r_2 y_1 \quad y_2 x_2 s_1 y_2 \quad y_3 s_2 x_3 t_1 y_3 \quad y_4 t_2 x_4 y_4 \\
& z \quad r_1 a_1 b_1 c_1 s_1 a_1 b_1 c_1 t_1 \quad r_2 a_2 b_2 c_2 s_2 a_2 b_2 c_2 t_2 \\
G_2 : \quad & r_1 x_1 y_1 x_1 r_2 \quad x_2 y_2 x_2 s_1 \quad s_2 x_3 y_3 x_3 t_1 \quad t_2 x_4 y_4 x_4 \\
& z \quad a_1 r_1 b_1 a_1 s_1 c_1 b_1 t_1 c_1 \quad a_2 r_2 b_2 a_2 s_2 c_2 b_2 t_2 c_2
\end{aligned}
$$

*The assignment $v_1 =$ true, $v_2 =$ false, $v_3 =$ false, $v_4 =$ true satisfies the 3SAT instance and corresponds to the following common reduced genome:*

$$G' : \quad r_1 x_1 y_1 \quad y_2 x_2 s_1 \quad y_3 x_3 t_1 \quad t_2 x_4 y_4 \quad z \quad a_1 b_1 c_1 \quad r_2 a_2 s_2 b_2 c_2$$

The reduction clearly runs in polynomial time. It remains to prove the following lemma:

**Lemma 1.** *The 3SAT instance $(V, E)$ is satisfiable if and only if the two genomes $G_1$ and $G_2$ have a common subsequence $G'$ including exactly one copy of each gene.*

We first prove the direct implication. Suppose that the 3SAT instance $(V, E)$ is satisfiable. We will compose a common subsequence $G'$ of the two genomes $G_1$ and $G_2$ from a common subsequence of each variable gadget $\langle v_i \rangle$, the separator gene $z$ in the middle, and a common subsequence of each clause gadget $\langle e_j \rangle$. Consider a truth assignment that satisfies the 3SAT instance. For each variable $v_i$, take the subsequence $p_{i,1} \ldots p_{i,k_i} x_i y_i$ if $v_i$ is set to true, and take the subsequence $y_i x_i q_{i,1} \ldots q_{i,l_i}$ if $v_i$ is set to false. For each clause $e_j$, at least one of its three literals is true; correspondingly, at least one of the three literal genes $r_j, s_j, t_j$ has been taken from some variable gadget $\langle v_i \rangle$. Now take a subsequence from the clause gadget $\langle e_j \rangle$ following one of three cases:

1. If $r_j$ has been taken, then take the subsequence $a_j b_j \underline{s_j} c_j \underline{t_j}$.

2. If $s_j$ has been taken, then take either the subsequence $\underline{r_j} b_j a_j c_j \underline{t_j}$ or the subsequence $\underline{r_j} a_j c_j b_j \underline{t_j}$.

3. If $t_j$ has been taken, then take the subsequence $\underline{r_j} a_j \underline{s_j} b_j c_j$.

Here an underlined literal gene is omitted from the subsequence taken from the clause gadget $\langle e_j \rangle$ if its other copy has already been taken from some variable gadget $\langle v_i \rangle$. The common subsequence $G'$ thus composed clearly includes exactly one copy of each gene.

We next prove the reverse implication. Suppose that the two genomes $G_1$ and $G_2$ have a common subsequence $G'$ including exactly one copy of each gene. We will find a satisfying assignment for the 3SAT instance $(V, E)$ as follows. Due to the strategic location of the separator gene $z$ in the two genomes, each literal gene must appear in the common subsequence either before $z$ in both genomes, in some variable gadget $\langle v_i \rangle$, or after $z$ in both genomes, in some clause gadget $\langle e_j \rangle$. The crucial property of the clause gadget $\langle e_j \rangle$ is that it cannot have a common subsequence including exactly one copy of each clause gene $a_j, b_j, c_j$ unless at least one of the three literal genes $r_j, s_j, t_j$ is omitted. A literal gene omitted from the common subsequence of the clause gadget $\langle e_j \rangle$ has to appear in the common subsequence of some variable gadget $\langle v_i \rangle$, where the two variable genes $x_i$ and $y_i$ must appear in the order $x_i y_i$ if the literal is positive and appear in the order $y_i x_i$ if the literal is negative. Now set each variable $v_i$ to true if the two variable genes $x_i$ and $y_i$ appear in the common subsequence $G'$ in the order $x_i y_i$, and set it to false otherwise. Then each clause gets at least one true literal. This completes the proof of Theorem 1.

## 3 Proof of Theorem 2

We prove that ZERO EXEMPLAR DISTANCE for multichromosomal genomes without gene order is NP-hard by a reduction again from 3SAT. Let $(V, E)$ be a 3SAT instance, where $V = \{v_1, \ldots, v_n\}$ is a set of $n$ boolean variables, $E = \{e_1, \ldots, e_m\}$ is a conjunctive boolean formula of $m$ clauses, and each clause in $E$ is a disjunction of exactly three literals of the variables in $V$. Without loss of generality, assume that no clause in $E$ contains two literals of the same variable in $V$. We will construct two genomes $G_1$ and $G_2$ over $n + 9m$ distinct genes:

- One *variable gene* $x_i$ for each variable $v_i$, $1 \le i \le n$;
- Six *clause genes* $a_j, b_j, c_j, a'_j, b'_j, c'_j$ for each clause $e_j$, $1 \le j \le m$;
- Three *literal genes* $r_j, s_j, t_j$ for the three literals of each clause $e_j$, $1 \le j \le m$.

For each variable $v_i$, let $p_{i,1}, \ldots, p_{i,k_i}$ be the literal genes for the positive literals of $v_i$, and let $q_{i,1}, \ldots, q_{i,l_i}$ be the literal genes for the negative literals of $v_i$. $G_1$ includes one subset and $G_2$ includes two subsets of genes including $x_i$:

$$G_1\langle v_i \rangle: \quad \{p_{i,1}, \ldots, p_{i,k_i}, x_i, q_{i,1}, \ldots, q_{i,l_i}\}$$
$$G_2\langle v_i \rangle: \quad \{p_{i,1}, \ldots, p_{i,k_i}, x_i\} \quad \{x_i, q_{i,1}, \ldots, q_{i,l_i}\}$$

For each clause $e_j$, $G_1$ includes six subsets and $G_2$ includes seven subsets of clause/literal genes:

$$G_1\langle e_j \rangle: \quad \{a_j, b_j\} \{b_j, c_j\} \{c_j, a_j\} \quad \{a'_j, r_j\} \{b'_j, s_j\} \{c'_j, t_j\}$$
$$G_2\langle e_j \rangle: \quad \{a_j, b_j, c_j\} \quad \{a_j, a'_j, r_j\} \{b_j, b'_j, s_j\} \{c_j, c'_j, t_j\} \quad \{a'_j\} \{b'_j\} \{c'_j\}$$

This completes the construction. It is easy to check that each gene appears at most two times in each genome, $G_1$ includes exactly $n + 15m$ genes including duplicates, and $G_2$ includes exactly $2n + 18m$ genes including duplicates. We give an example:

**Example 2.** *For a 3SAT instance of 4 variables and 2 clauses $e_1 = \{r_1 = v_1, s_1 = \neg v_2, t_1 = \neg v_3\}$ and $e_2 = \{r_2 = \neg v_1, s_2 = v_3, t_2 = v_4\}$, the reduction constructs the following two genomes:*

$$\begin{aligned}
G_1: \quad & \{r_1, x_1, r_2\} \quad \{x_2, s_1\} \quad \{s_2, x_3, t_1\} \quad \{t_2, x_4\} \\
& \{a_1, b_1\} \{b_1, c_1\} \{c_1, a_1\} \quad \{a'_1, r_1\} \{b'_1, s_1\} \{c'_1, t_1\} \\
& \{a_2, b_2\} \{b_2, c_2\} \{c_2, a_2\} \quad \{a'_2, r_2\} \{b'_2, s_2\} \{c'_2, t_2\} \\
G_2: \quad & \{r_1, x_1\} \{x_1, r_2\} \quad \{x_2\} \{x_2, s_1\} \quad \{s_2, x_3\} \{x_3, t_1\} \quad \{t_2, x_4\} \{x_4\} \\
& \{a_1, b_1, c_1\} \quad \{a_1, a'_1, r_1\} \{b_1, b'_1, s_1\} \{c_1, c'_1, t_1\} \quad \{a'_1\} \{b'_1\} \{c'_1\} \\
& \{a_2, b_2, c_2\} \quad \{a_2, a'_2, r_2\} \{b_2, b'_2, s_2\} \{c_2, c'_2, t_2\} \quad \{a'_2\} \{b'_2\} \{c'_2\}
\end{aligned}$$

*The assignment $v_1 = \text{true}, v_2 = \text{false}, v_3 = \text{false}, v_4 = \text{true}$ satisfies the 3SAT instance and corresponds to the following common reduced genome:*

$$G' : \quad \begin{array}{llll} \{r_1, x_1\} & \{x_2, s_1\} & \{x_3, t_1\} & \{t_2, x_4\} \\ \{a_1\}\,\{b_1, c_1\} & \{a'_1\}\,\{b'_1\}\,\{c'_1\} & & \\ \{c_2\}\,\{a_2, b_2\} & \{a'_2, r_2\}\,\{b'_2, s_2\}\,\{c'_2\} & & \end{array}$$

The reduction clearly runs in polynomial time. It remains to prove the following lemma:

**Lemma 2.** *The 3SAT instance $(V, E)$ is satisfiable if and only if the two genomes $G_1$ and $G_2$ have a common reduced genome $G'$ including exactly one copy of each gene.*

We first prove the direct implication. Suppose that the 3SAT instance $(V, E)$ is satisfiable. We will compose a common reduced genome $G'$ of the two genomes $G_1$ and $G_2$ as follows. Consider a truth assignment that satisfies the 3SAT instance. For each variable $v_i$, take the subset $\{p_{i,1}, \ldots, p_{i,k_i}, x_i\}$ if $v_i$ is set to true, and take the subset $\{x_i, q_{i,1}, \ldots, q_{i,l_i}\}$ if $v_i$ is set to false. For each clause $e_j$, at least one of its three literals is true; correspondingly, at least one of the three literal genes $r_j, s_j, t_j$ has been taken from some variable gadget $\langle v_i \rangle$. Now take some subsets of clause/literal genes following one of three cases:

1. If $r_j$ has been taken, then take the subsets $\{a_j\}, \{b_j, c_j\}, \{a'_j\}, \{b'_j, \underline{s_j}\}, \{c'_j, \underline{t_j}\}$.

2. If $s_j$ has been taken, then take the subsets $\{b_j\}, \{c_j, a_j\}, \{a'_j, \underline{r_j}\}, \{b'_j\}, \{c'_j, \underline{t_j}\}$.

3. If $t_j$ has been taken, then take the subsets $\{c_j\}, \{a_j, b_j\}, \{a'_j, \underline{r_j}\}, \{b'_j, \underline{s_j}\}, \{c'_j\}$.

Here an underlined literal gene is omitted from the subset taken from the clause gadget $\langle e_j \rangle$ if its other copy has already been taken from some variable gadget $\langle v_i \rangle$. The reduced genome $G'$ thus composed clearly includes exactly one copy of each gene.

We next prove the reverse implication. Suppose that the two genomes $G_1$ and $G_2$ have a common reduced genome $G'$ including exactly one copy of each gene. We will find a satisfying assignment for the 3SAT instance $(V, E)$ as follows. The crucial property of the clause gadget $\langle e_j \rangle$ is that it cannot have a common reduced genome including exactly one copy of each clause gene $a_j, b_j, c_j, a'_j, b'_j, c'_j$ unless at least one of the three literal genes $r_j, s_j, t_j$ is omitted. A literal gene omitted from the clause gadget $\langle e_j \rangle$ has to appear in a subset in $G'$ that contains some variable gene $x_i$. By the construction of the variable gadgets, this subset contains, besides $x_i$, either literal genes for positive literals, or literal genes for negative literals. Now set each variable $v_i$ to true if the subset in $G'$ that contains $x_i$ also contains at least one literal gene for a positive literal, and set it to false otherwise. Then each clause gets at least one true literal. This completes the proof of Theorem 2.

## 4 Proof of Theorem 3

Let $A$ and $B$ be two sequences of lengths $n$ and $m$, respectively, over an alphabet $\Sigma = \Sigma_1 \cup \Sigma_2$, where $\Sigma_1$ is a set of mandatory symbols and $\Sigma_2$ is a set of optional symbols. In the special case that each mandatory symbol in $\Sigma_1$ appears exactly once in one sequence and at least once in the other sequence, we have the obvious but important property that *any common subsequence of the two sequences can contain each mandatory symbol at most once*. This property leads to a very simple algorithm that decides whether a feasible solution to EXEMPLAR LONGEST COMMON SUBSEQUENCE exists in this special case:

**Algorithm 1.**

1. Obtain two sequences $A'$ and $B'$ from $A$ and $B$ by deleting all optional symbols in $\Sigma_2$.

2. Compute a longest common subsequence $C^*$ of $A'$ and $B'$.

3. If $C^*$ contains all mandatory symbols in $\Sigma_1$, return yes. Otherwise, return no.

The time complexity of Algorithm 1 is $O(nm)$ by using a standard dynamic programming algorithm for longest common subsequence [10]. The correctness of Algorithm 1 is justified by the following lemma:

**Lemma 3.** *A and B have a common subsequence containing all mandatory symbols in $\Sigma_1$ if and only if the longest common subsequence $C^*$ of $A'$ and $B'$ contains all mandatory symbols in $\Sigma_1$.*

*Proof.* The reduction from $A$ and $B$ to $A'$ and $B'$ preserves the mandatory symbols. Thus $A$ and $B$ have a common subsequence containing all mandatory symbols in $\Sigma_1$ if and only if $A'$ and $B'$ have a common subsequence containing all mandatory symbols in $\Sigma_1$. It remains to prove the equivalent claim that $A'$ and $B'$ have a common subsequence containing all mandatory symbols in $\Sigma_1$ if and only if $C^*$ contains all mandatory symbols in $\Sigma_1$.

The "if" direction of the claim is trivial because $C^*$ is a common subsequence of $A'$ and $B'$. To prove the "only if" direction, recall that in any common subsequence of $A'$ and $B'$, each mandatory symbol can appear at most once. Thus the length of any common subsequence of $A'$ and $B'$ is at most the size of $\Sigma_1$. Moreover, if the length of some common subsequence of $A'$ and $B'$ is equal to the size of $\Sigma_1$, then this common subsequence must contain all mandatory symbols in $\Sigma_1$, and vice versa. Now suppose that $A'$ and $B'$ have a common subsequence $C'$ containing all mandatory symbols in $\Sigma_1$. Then the length of $C'$ must be equal to the size of $\Sigma_1$. Since the length of $C^*$ is at least the length of $C'$, the length of $C^*$ must also be equal to the size of $\Sigma_1$. Then $C^*$ must contain all mandatory symbols in $\Sigma_1$ too. This completes the proof. □

Since deciding whether a feasible solution to EXEMPLAR LONGEST COMMON SUBSEQUENCE exists for two sequences $A$ and $B$ is the same as the problem ZERO EXEMPLAR DISTANCE for two monochromosomal genomes $A'$ and $B'$ obtained from $A$ and $B$ by deleting all optional symbols, we also have an $O(nm)$ algorithm for ZERO EXEMPLAR DISTANCE for monochromosomal genomes in the special case that each gene appears exactly once in one genome and at least once in the other genome.

We next present an algorithm for the maximization problem EXEMPLAR LONGEST COMMON SUBSEQUENCE in the special case that each mandatory symbol appears exactly once in one input sequence and at least once in the other input sequence:

**Algorithm 2.**

1. Assign each mandatory symbol in $\Sigma_1$ a weight of $w = \min\{n, m\} + 1$, and assign each optional symbol in $\Sigma_2$ a weight of 1. Compute a common subsequence $C^*$ of $A$ and $B$ of the maximum total weight.

2. If $C^*$ contains all mandatory symbols in $\Sigma_1$, return $C^*$. Otherwise, report that no feasible solution exists.

If $A$ and $B$ have no common subsequence containing all mandatory symbols in $\Sigma_1$, then clearly the maximum-weight common subsequence $C^*$ of $A$ and $B$ cannot contain all mandatory symbols in $\Sigma_1$, and hence the algorithm correctly reports that no feasible solution exists. Otherwise, the correctness of Algorithm 2 is justified by the following lemma:

**Lemma 4.** *If $A$ and $B$ have a common subsequence containing all mandatory symbols in $\Sigma_1$, then the maximum-weight common subsequence $C^*$ of $A$ and $B$ is a longest common subsequence of $A$ and $B$ that contains all mandatory symbols in $\Sigma_1$.*

*Proof.* Suppose that $A$ and $B$ have a common subsequence $C$ containing all mandatory symbols in $\Sigma_1$. We first show that the maximum-weight common subsequence $C^*$ of $A$ and $B$ contains all mandatory symbols in $\Sigma_1$. Note that the number of optional symbols in $C^*$ is at most the length of $C^*$, which is at most $\min\{n, m\}$. Also recall that any common subsequence of $A$ and $B$ can contain each mandatory symbol at most once. If $C^*$ does not contain all mandatory symbols in $\Sigma_1$, then by our choice of $w = \min\{n, m\} + 1$, the total weight of $C^*$ would be at most

$$(|\Sigma_1| - 1) \cdot w + \min\{n, m\} \cdot 1 < (|\Sigma_1| - 1) \cdot w + w \cdot 1 = |\Sigma_1| \cdot w.$$

On the other hand, since $C$ contains all mandatory symbols in $\Sigma_1$, the weight of $C$ is at least $|\Sigma_1| \cdot w$. This contradicts the assumption that $C^*$ is a maximum-weight common subsequence of $A$ and $B$.

Now, since $C^*$ contains all mandatory symbols and can contain each mandatory symbol at most once, $C^*$ must contain each mandatory symbol exactly once. Then, to have the maximum total weight, $C^*$ must be a longest common subsequence of $A$ and $B$ that contains all mandatory symbols in $\Sigma_1$. □

Again, the overall time complexity of Algorithm 2 is clearly $O(nm)$. This completes the proof of Theorem 3.

## 5 Proof of Theorem 4

We present two algorithms for ZERO EXEMPLAR DISTANCE for multichromosomal genomes without gene order. Let $k_1$ and $k_2$, respectively, be the numbers of chromosomes in $G_1$ and $G_2$. Let $A_1, \ldots, A_{k_1}$ be the $k_1$ chromosomes in $G_1$. Let $B_1, \ldots, B_{k_2}$ be the $k_2$ chromosomes in $G_2$. Let $k = \max\{k_1, k_2\}$. Let $n$ be the total number of genes in $G_1$ and $G_2$, i.e., $n = \sum_{i=1}^{k_1} |A_i| + \sum_{j=1}^{k_2} |B_j|$.

We first present a polynomial-time algorithm for ZERO EXEMPLAR DISTANCE for multichromosomal genomes without gene order in the special case that each gene appears exactly once in one genome and at least once in the other genome. Our algorithm is based on maximum-weight matching in bipartite graphs:

**Algorithm 3.**

1. Construct a complete bipartite graph $G = (V_1 \cup V_2, V_1 \times V_2)$ with vertices $V_1 = \{A_1, \ldots, A_{k_1}\}$ and $V_2 = \{B_1, \ldots, B_{k_2}\}$. Associate with each edge between $A_i \in V_1$ and $B_j \in V_2$ a reduced chromosome $C_{ij} = A_i \cap B_j$ and a weight equal to its size.

2. Compute a maximum-weight matching $M$ in the graph $G$.

3. If the set of reduced chromosomes corresponding to the edges in $M$ includes all the genes, return yes. Otherwise, return no.

To see the correctness of Algorithm 3, note that each reduced chromosome of a common reduced genome is a common subset of two distinct chromosomes, one from each input genome, and corresponds to an edge of a matching in the complete bipartite graph. In the special case that each gene appears exactly once in one genome and at least once in the other genome, no gene can appear more than once in the reduced chromosomes corresponding to the edges of a matching. Thus the maximum possible weight of a matching is equal to the number of distinct genes, and a common reduced genome that includes all the genes corresponds to a matching of the maximum weight.

We now analyze the time complexity of Algorithm 3. Steps 1 and 3 can be easily implemented in $O(n^2)$ time. Step 2 can be implemented in $O(k^3)$ time using a standard algorithm for weight bipartite matching; see e.g. [13]. Thus the overall time complexity is $O(n^2 + k^3)$.

We next present a fixed-parameter tractable algorithm for this problem without any assumption on the distribution of duplicate genes. Refer to [7] for basic concepts in parameterized complexity theory. The parameter of our algorithm is $k = \max\{k_1, k_2\}$:

**Algorithm 4.**

1. Add $k - k_1$ empty chromosomes $A_{k_1+1}, \ldots, A_k$ to $G_1$, or add $k - k_2$ empty chromosomes $B_{k_2+1}, \ldots, B_k$ to $G_2$, such that $G_1$ and $G_2$ have the same number $k$ of chromosomes.

2. For each permutation $\pi$ of $\langle 1, \ldots, k \rangle$, compute $C_\pi = \cup_{i=1}^{k}(A_i \cap B_{\pi(i)})$.

3. If for some permutation $\pi$ the set $C_\pi$ includes all the genes, return yes. Otherwise return no.

To see the correctness of Algorithm 4, note again that each chromosome of a common reduced genome is a common subset of two distinct chromosomes, one from each input genome. All other chromosomes of the two input genomes that do not contribute to the common reduced genome are deleted. To handle the matching and the deletion of the chromosomes in a uniform way, we can think of each chromosome deleted from one genome as matched to a chromosome deleted from the other genome or to an empty chromosome. Thus by padding the two genomes to the same number of chromosomes, we only need to consider perfect matchings as permutations. The time complexity of Algorithm 4 is $O(k!\, n^2)$, with $O(n^2)$ time for each of the $k!$ permutations. This completes the proof of Theorem 4.

We remark that the problem ZERO EXEMPLAR DISTANCE for multichromosomal genomes without gene order is unlikely to have a fixed-parameter tractable algorithm if the parameter is the maximum number of genes in any single chromosome. This is because 3SAT remains NP-hard even if for each variable there are at most five clauses that contain its literals [9]. As a result, the number of genes in each chromosome need not be more than some constant in our reduction from 3SAT.

**Acknowledgment** The author would like to thank Binhai Zhu for a brief discussion on Question 1 during a visit to University of Texas – Pan American in May 2008, and thank Guillaume Fertin for communicating the recent result [3]. The alternative proof of Theorem 1 was obtained independently by the author in February 2010 without knowing the recent progress [3]. The author also thanks Pedro J. Tejada for bringing the open question of Bonizzoni et al. [4], Question 2, to his attention.